\documentclass[12pt]{article}
\usepackage[english,activeacute]{babel}
\usepackage{natbib}
\usepackage{comment}
\usepackage{float}
\usepackage[hidelinks]{hyperref}
\usepackage{mathrsfs}
\usepackage{enumitem}
\usepackage[font={small,it}]{caption}
\usepackage{amsmath,amsfonts,amsthm,amssymb}
\usepackage{bm,rotating,multirow,dsfont,graphicx}
\usepackage[usenames, dvipsnames]{color}
\usepackage{url}
\usepackage{multicol}
\usepackage{multirow}
\usepackage[T1]{fontenc}
\usepackage{flafter}
\usepackage{appendix}
\usepackage{subfigure}
\usepackage{xcolor}
\makeatletter
\def\hlinewd#1{%
	\noalign{\ifnum0=`}\fi\hrule \@height #1 %
	\futurelet\reserved@a\@xhline}
\makeatother
\def\@roman#1{\romannumeral #1}

\begin{document}

\def\spacingset#1{\renewcommand{\baselinestretch}{#1}\small\normalsize}\spacingset{1}

\title{Bayesian spatial voting model to characterize the legislative behavior of the Colombian Senate 2010--2014}

\date{}
	
\author{
    Carolina Luque, Universidad Ean, Colombia\footnote{cluque2.d@universidadean.edu.co} \\ 
    Juan Sosa, Universidad Nacional, Colombia\footnote{jcsosam@unal.edu.co}
}	 

\maketitle

\begin{abstract} 
This paper applies Bayesian methodologies to characterize the legislative behavior of the Colombian Senate during the period 2010--2014. The analysis is carried out through the plenary roll call votes of this legislative chamber. In addition, parliamentary electoral behavior is operationalized by implementing the one-dimensional standard Bayesian ideal point estimator via the Markov chain Monte Carlo algorithms. The results contribute mainly to two points: political space dimensionality and the identification of pivot legislators. The pattern revealed by the estimated ideal points suggests a latent non--ideological trait (opposition - non--opposition) underlying the vote of deputies in the Senate. Thus, in addition to providing empirical evidence for a better understanding of legislative policy in Colombia during the period under analysis, this work also offers methodological and theoretical tools to guide the analysis of roll call vote data in contexts of unbalanced parliaments (as opposed to the U.S. parliament), taking the particular case of the Colombian Senate as a reference.
\end{abstract}


\spacingset{1.1} 

\section{Introduction}

The spatial theory of voting establishes theoretical and methodological considerations under which spatial voting models are developed. These models bring together a set of mathematical and statistical tools that allow the study of the competitive behavior of political agents. Their implementation has played a prominent role in political science from the 1950s (\citealp{krehbiel1988spatial}; \citealp{arrow1990advances}). Ever since then, many publications have increased the visibility and highlighted the usefulness of these models to provide empirical evidence that supports the theoretical explanation of the electoral behavior and the forms of electoral organization in different contexts. Among the spatial voting models that stand out, the Bayesian ideal point estimator (\citealp{clinton2004statistical}; \citealp{jackman2004bayesian}), whose specification and implementation dates back to the last decade, is to be named. Literature shows that this model is a fundamental tool for research in Political Science \citep{yu2019circle}, and also warns that it is a dominant methodological and theoretical tool for analyzing modern voting patterns (\citealt{yu2019spherical}; \citealt{moser2019multiple}).

Studies on legislative behavior in Colombia point to an important field of research (see \citealp{archer1997unr}; \citealp{cardenas2008political}; \citealp{aleman2008comisiones}; \citealp{pachon2011que}; \citealp{carroll2016unrealized}; \citealp{pachon2016s}; \citealp{pachon2020}, among others). An exhaustive literature review shows that quantitative research about the Colombian parliament has been mainly directed towards the search for patterns and determinants of legislative behavior at the party, coalition, and committee levels (e.g., \citealt{carroll2016unrealized}; \citealt{pachon2016s}), typically leaving the political behavior of each legislator behind.

Analysis at the level of the individual is quite relevant in a context such as Colombia, given that the country's political system is characterized by a personalized policy (\citealt{pachon2011que}; \citealt{pachon2020}). Additionally, most studies on legislative behavior support their results through descriptive analyses of bills that have been introduced and voted in Congress (e.g., \citealt{pachon2011que}; \citealt{pachon2020}), showing an opportunity to make estimates and inferences on legislative electoral behavior in \textit{unbalanced parliaments} (multi--party and composed of factions that do not have an equivalent number of seats) such as the Colombian one.

Few studies apply any spatial voting model to analyze parliamentary voting behavior. To the best of our knowledge, there is only one work \citep{carroll2016unrealized} that applies the Aldrich--McKelvey scaling method \citep{hare2015using} and optimal classification \citep{poole2000nonparametric} to account for patterns of legislative electoral behavior by means of survey data and nominal votes. Specifically, no research that uses ideal points to characterize the political preferences of Colombian legislators from nominal voting data are available.

Along this road, we propose the implementation of Bayesian methods to characterize the legislative behavior of the Colombian Senate 2010--2014. Specifically, we seek not only to recognize the conditions under which the results achieved are invariant when applying the one-dimensional standard Bayesian ideal point estimator to the nominal voting data, but also to distinguish the latent features, and the partisan and coalition patterns that underlie the legislative decisions of the deputies, in addition to analyzing the roll call vote data. In order to evaluate the model's performance, we conducted several experiments using synthetic data considering different configurations of anchor legislators, missing data rates (abstentions and nonattendances), link functions, and prior information, under the context of an unbalanced parliament.

This document is structured as follows: Section \ref{sec:colombia} presents a synthesis of the quantitative studies on legislative behavior in Colombia; Section \ref{sec:Modelo} shows the theoretical aspects of the Bayesian ideal point estimator, together with the particularities of the dimension of the political space and the identification of pivot legislators; Section \ref{sec:hiperYcomputo} extends the details of implementation of the one--dimensional standard Bayesian ideal point estimator; Section \ref{sec:simulacion} states the results of the simulation study; Section \ref{sec:apli} analyzes the data for the Colombian Senate 2010--2014; and finally, Section \ref{sec:con} discusses the results of the work and some alternatives for future research.

\section{Quantitative studies on legislative behavior in Colombia} \label{sec:colombia}

Studies that apply a spatial voting model to analyze parliamentary electoral behavior in the Colombian context are scarce. In particular, to date, there is no research that implements the Bayesian ideal point estimator to estimate individual preferences of legislators from nominal voting data (\citealp{clinton2004statistical}; \citealp{jackman2004bayesian}).

In the Colombian case, political studies that account for the implementation of any spatial voting model date back to the 1970s (e.g., \citealp{hoskin1973inter}). Research from that period was oriented to the analysis of partisan behavior in general terms, without focusing on the study of the individual preferences of the members of a specific institution, such as the parliament. The results of these studies are based on surveys and allude to the preferences of political leaders and groups. In addition, they raise initial discussions about the dimension of political space from a partisan point of view (e.g., \citealp{hoskin1973inter}; \citealp{hoskin1979belief}). For example, \cite{hoskin1973inter}, apply a version of the spatial model that is presented in \cite{downs1957economic} based on surveys to political actors with different hierarchies and from different parties. Such implementation leads to identify two outstanding dimensions in the conflict of the Colombian party system, an opposition--government pole and a left--right ideological pole. The results achieved by these authors point out that the conflict centered on the control of the government is more forceful than the ideological conflict. None of current studies related to the dimension of political space provide conclusions in terms of the dimension of political space in the particular scenario of the Colombian parliament.

The work directed specifically at the analysis of the legislative chambers in Colombia also dates back to the 1970s (\citealp{kline1974interest}; \citealp{hoskin1975dimensions}; \citealp{hoskin1976legislative}; \citealp{kline1977committee}). These studies characterize the legislature from a descriptive point of view (e.g., \citealp{kline1977committee}), and in a context prior to the 1991 constitutional reform, which substantially modifies the regulation of the Congress of the Republic and the policy-making process in the country (\citealp{pachon2020}; \citealp{cardenas2008political}). These studies do not allude to the use of voting records for the analysis of electoral behavior within legislative bodies. The usefulness of voting data for the characterization of parliamentary electoral behavior in Colombia is recent, since they have been collected as of 2006 \citep{carroll2016unrealized}.

After the reform, a contemporary field of research in relation to quantitative studies on legislative behavior was identified. Studies such as the ones developed by \cite{archer1997unr}, \cite{cardenas2008political}, \cite{aleman2008comisiones}, \cite{pachon2011que}, \cite{carroll2016unrealized}, \cite{pachon2016s}, \cite{pachon2020}, among others, offer a look at the Colombian legislature from its current context. Some of them provide empirical evidence regarding the programmatic nature, nominal voting patterns, and legislative activity trends of political groups (\citealp{carroll2016unrealized}; \citealp{pachon2016s}).

The quantitative analysis of parliamentary behavior in Colombia is directed towards the search for patterns and determinants of legislative behavior, predominantly at the party, coalition, and committee levels; leaving the field of research at the legislator level open, which is of vital importance given the existence of a personalized policy in the country (\citealp{pachon2011que}; \citealp{pachon2020}). Several studies base their findings on the descriptive analyses of bills that have been deliberated in both legislative chambers (Senate and House of Representatives; e.g., \citealp{pachon2020}; \citealp{carroll2016unrealized}; \citealp{pachon2011que}). For example, by means of surveys and congressional voting reports, \cite{pachon2020} identify percentage patterns in the proportion of bills introduced by each committee in both chambers, and regularities in constitutional amendments passed and defeated in committee and plenary between 1998 and 2018.

Other more sophisticated statistical techniques have been used to analyze Colombian legislative behavior. \cite{pachon2016s} propose frequentist logistic regression models with robust standard errors \citep{croux2004robust} to determine the factors that push the committee chairperson to choose members from his own party or coalition, such as rapporteurs, and the relevant factors for a rapporteur to give a positive or negative report to a bill in the first debate. On the other hand, \cite{carroll2016unrealized} analyze legislative voting patterns through frequentist probit regression models in order to examine the bills’ approval rate considering the initiative (executive or legislative) and the scope (national or local) of these bills. In addition, they use data from the Parliamentary Elites Latin America Project (PELA) from the period 1998--2010 to implement a Bayesian version of the Aldrich--McKelvey scaling method \citep{hare2015using} in a one--dimensional (left--right) space, and to examine the distribution of ideological preferences (self--reported) of the parties comprising the Colombian House of Representatives. Finally, they also use nominal voting records to examine aggregate voting patterns (by party) in Congress during 2006--2014. In this case, they apply an optimal classification method \citep{poole2000nonparametric}, with which they estimate the political preferences of the parties with the largest number of seats in Congress. The results show partisan tendencies, and their conclusions do not allude to individual political preferences. The literature does not note studies focused exclusively on the Senate of the Republic of Colombia.

\section{Bayesian Spatial Voting Modeling} \label{sec:Modelo}

Roll call data are generated when $n$ legislators vote on $m$ motions (bills, legislative initiatives, etc.); each legislator $i \in \{1, \ldots, n \}$ must take a position in favor of or against motion $j \in \{1, \ldots, m \}$. Operationalizing the legislative voting behavior, the votes in favor of or against motion $j$ are conceptualized as points in a $d$-dimensional Euclidean space (known as political space), and denoted by $\boldsymbol {\psi}_{j}$ and $\boldsymbol{\zeta}_{j}$, respectively.

Let $y_{i,j} \in \{0,1\}$ be the vote cast by legislator $i$ on motion $j$, with $y_{i,j}=1$ if such a vote turns out in favor of the motion, and $y_{i,j}=0$ otherwise. Also, it is assumed that all the legislators have a political preference, i.e., each legislator $i$ has a latent (unobserved) factor $\boldsymbol{\beta}_{i} \in \mathbb{R}^d$ known as ``ideal point''. Thus, considering both voting alternatives and ideal points, decisions are made based on a quadratic utility function defined over the political space given by
\begin{equation} \label{equ:utility}
    U_i(\boldsymbol{\psi}_{j})=-\parallel\boldsymbol{\psi}_{j}-\boldsymbol{\beta}_{i}\parallel^2+\eta_{i,j} 
    \qquad\text{and}\qquad
    U_i(\boldsymbol{\zeta}_{j})=-\parallel\boldsymbol{\zeta}_{j}-\boldsymbol{\beta}_{i}\parallel^2+\upsilon_{i,j}
\end{equation}
where $U_i(\boldsymbol{\psi}_{j})$ and $U_i(\boldsymbol{\zeta}_{j})$ are the corresponding profits associated with legislator $i$ voting in favor of or against motion $j$, respectively, $\parallel \cdot \parallel$ is the Euclidean norm in $\mathbb{R}^d$, and finally, $\eta_{i,j}$ and $\upsilon_{i,j}$ are independent stochastic deviations (random shocks) resulting from the uncertainty involved in the voting processes, whose joint probabilistic distribution is such that $\textsf{E}(\eta_{i,j} -\upsilon_{i,j})=0$ and $\textsf{Var} (\eta_{i,j} - \upsilon_{i,j}) = \sigma^2_{j}$. Even though there are exist other meaningful ways to define both the political space and the utility function, they fall outside the scope of this this work and will be investigated elsewhere.

Rational choice theory (e.g., \citealp{clinton2004statistical}; \citealp{yu2019spherical}) states that under the previous setting, the legislator $i$ votes in favor of motion $j$ if and only if $U_i(\boldsymbol{\psi}_{j})> U_i(\boldsymbol{\zeta}_{j})$, i.e.,
\begin{equation*}
y_{i,j} \mid  \boldsymbol{\zeta}_{j},\boldsymbol{\psi}_{j}, \sigma_{j}, \boldsymbol{\beta}_{i} =
    \begin{cases}
    1, & \text{si $U_i(\boldsymbol{\psi}_{j}) - U_i(\boldsymbol{\zeta}_{j})$} > 0 \\
    0, & \text{otherwise}
    \end{cases}
\end{equation*}
and therefore,
$$
\mathrm{Pr}(y_{i,j}=1 \mid \boldsymbol{\zeta}_{j},\boldsymbol{\psi}_{j}, \sigma_{j}, \boldsymbol{\beta}_{i}) = \mathrm{Pr}(\epsilon_{i,j} < \mu_{j}+\boldsymbol{\alpha}_{j}^{\textsf{T}}\boldsymbol{\beta}_{i})
= G(\mu_{j}+\boldsymbol{\alpha}_{j}^{\textsf{T}}\boldsymbol{\beta}_{i})
$$  
where $\epsilon_{i,j}=(\upsilon_{i,j}-\eta_{i,j})/ \sigma_j$ is the random quantity associated with the voting process,  
$\mu_{j}=(\boldsymbol{\zeta}_{j}^{\textsf{T}} \boldsymbol{\zeta}_{j}-\boldsymbol{\psi}_{j}^{\textsf{T}}\boldsymbol{\psi}_{j})/ \sigma_{j}$ is the ``approval'' parameter representing the basal probability of a vote in favor of motion $j$, $\boldsymbol{\alpha}_{j}=2(\boldsymbol{\psi}_{j}-\boldsymbol{\zeta}_{j})/ \sigma_{j}$ is the ``discrimination'' parameter representing the effect of the ideal points on the probability of a vote in favor of motion $j$ across the political space dimensions, and finally, $G(\cdot)$ is an appropriate link function. If $\epsilon_{i,j}\sim\textsf{N}(0,1)$, then $G$ is a probit link, i.e., $G(\mu_j+\boldsymbol{\alpha}_j^{\textsf{T}}\boldsymbol{\beta}_i)=\boldsymbol{\Phi}(\mu_j+\boldsymbol{\alpha}_j^{\textsf{T}}\boldsymbol{\beta}_i)$, where $\boldsymbol{\Phi}(\cdot)$ is the cumulative distribution function of the standard Normal distribution. On the other hand, if $\epsilon_{i,j}$ follows a Standard Logistic distribution, then $G$ is a logit link, i.e., $G(\mu_j+\boldsymbol{\alpha}_j^{\textsf{T}}\boldsymbol{\beta}_i) = \mathrm{expit}(\mu_{j}+\boldsymbol{\alpha}_{j}^{\textsf{T}}\boldsymbol{\beta}_{i})$, where $\mathrm{expit}(x) = 1/(1+\exp{(-x)})$.

The previous formulation leads to a latent factor model for binary data, which fully characterizes the probability of obtaining a positive vote since
\begin{equation*}
  y_{i,j} \mid \mu_j,\boldsymbol{\alpha}_j,\boldsymbol{\beta}_i \stackrel{\text {iid}}{\sim} \text{Ber}( G(\mu_j+\boldsymbol{\alpha}^{\textsf{T}}_{j}\boldsymbol{\beta}_i))  
\end{equation*}
and therefore, the likelihood of the model is
\begin{equation}\label{eqn:ModeloBase}
    p(\mathbf{Y} \mid \boldsymbol{\mu},\boldsymbol{A},\boldsymbol{B})=\prod_{i=1}^{n}\prod_{j=1}^{m}G(\mu_{j}+\boldsymbol{\alpha}_j^{\textsf{T}}\boldsymbol{\beta}_{i})^{y_{i,j}}\left [ 1-G(\mu_{j}+\boldsymbol{\alpha}_j^{\textsf{T}}\boldsymbol{\beta}_{i})\right]^{1-y_{i,j}}
\end{equation}
where $\mathbf{Y}=[y_ {i, j}]$ is a binary matrix of size $n \times m$, $\boldsymbol{\mu}=(\mu_1, \cdots, \mu_m)$ is a column vector of size $d\times 1$, and $\boldsymbol{A}=[\boldsymbol{\alpha}_{1}, \ldots, \boldsymbol {\alpha}_{m}]^{\textsf{T}}$ and $\boldsymbol{B}=[\boldsymbol{\beta}_{1}, \ldots, \boldsymbol{\beta}_{n}]^{\textsf{T}}$ are rectangular matrices of size $m \times d$ and $n \times d$, respectively.

In order to carry out a fully Bayesian specification of the model, it is mandatory to specify a joint prior distribution on $(\boldsymbol{\mu},\boldsymbol{A},\boldsymbol{B})$. A simple but powerful alternative that works well in practice relies on assuming Normal independent priors, which leads to a semiconjugate analysis and higher computational efficiency (\citealp{jackman2004bayesian}; \citealp{yu2019circle}; \citealp{yu2019spherical}). Thus, it follows that 
\begin{equation}\label{eqn:PreviaIdeal}
    (\mu_{j},\boldsymbol{\alpha}_j) \mid \boldsymbol{a}_{0}, \boldsymbol{A}_{0} \stackrel{\text {iid}}{\sim} \text{N}( \boldsymbol{a}_{0}, \boldsymbol{A}_{0})
    \qquad\text{and}\qquad
   \boldsymbol{\beta}_i \mid \boldsymbol{b}_{i}, \boldsymbol{B}_{i} \stackrel{\text {ind}}{\sim} \text{N}(\boldsymbol{b}_{i}, \boldsymbol{B}_{i})
\end{equation}
where $\boldsymbol{a}_0, \boldsymbol{A}_0, \boldsymbol{b}_i$, and $\boldsymbol{B}_i$ are the hyperparameters of the model. Note that $\boldsymbol{a}_{0}$ and $\boldsymbol{b}_{i}$ are mean column vectors of size $(d + 1)\times 1$ and $d\times 1$, respectively, and $\boldsymbol{A}_{0}$ and $\boldsymbol{B}_{i}$ are covariance matrices of size $(d + 1) \times (d + 1)$ and $d \times d$, respectively. As a final remark, some authors \citep[e.g.,][]{clinton2004statistical} propose setting $\boldsymbol{a}_{0}= \boldsymbol{0}_{(d + 1)}$ and $\boldsymbol{A}_{0}=\sigma^2\mathbf{I}_{(d + 1)}$ with $\sigma^2$ an arbitrarily large constant in order to assign a zero-centered non-informative prior to $\mu_j$ and $\boldsymbol{\alpha}_j$, as well as setting $\boldsymbol{b}_{i}=\boldsymbol{0}_d$ and $\boldsymbol{B}_{i}=\mathbf{I}_{d}$ for each $\boldsymbol{\beta}_i$, which imposes a identification restriction on the ideal points based on the prior distribution (see Section \ref{sec:identi} for more details).

\subsection{Posterior Inference}\label{sec:estim_infer}

a $d$-dimensional Euclidean spatial voting model involves many parameters. Specifically, considering data from $n$ legislators on $m$ voting proposals, the model requires $dn+m(d+1)$ parameters, out of which $nd$ correspond to ideal points, whereas $m(d+1)$ to voting-proposal-specific parameters. Such a multiparameter space makes the posterior distribution $p(\boldsymbol{\mu}, \boldsymbol{A}, \boldsymbol{B}\mid \mathbf{Y})$ particularly difficult to obtain since it is a distribution framed in a high dimension and analytically intractable.

We consider Markov chain Monte Carlo algorithms (MCMC; e.g., \citealp{gamerman2006markov}) to approximate the posterior distribution, since the prior specification allows us to structure tractable simulation-based algorithms. In particular, the Gibbs sampler ensures that a sequence of dependent bu approximately independent draws from the posterior distribution can be generated, through iterative sampling from the full conditional distributions of $\mu_{j}, \boldsymbol{\alpha}_{j}$ and $\boldsymbol{\beta}_{i}$. Thus, point and interval estimates can be approximated from the corresponding empirical distributions. Details about the MCMC algorithms implemented here can be found in Appendix \ref{appendix}.

\subsection{Identifiability}\label{sec:identi}

The model parameters in \eqref{eqn:ModeloBase} and \eqref{eqn:PreviaIdeal} are not identifiable. Specifically, note that Euclidean distances among ideal points $\boldsymbol{\beta}_i$ and the voting alternatives $\boldsymbol{\psi}_{j}$ and $\boldsymbol{\zeta}_{j}$ remain invariant under any change of scale, translation, rotation, or reflection of the political space. Such a geometric occurrence ensures that both discrimination parameters and ideal points are not distinguishable for any voting pattern $\mathbf{Y}$ (\citealp{clinton2004statistical}; \citealp{jackman2004bayesian}). For instance, consider a rotation of the political space through an $d \times d$ orthogonal matrix $\mathbf{Q}$ (i.e., $\mathbf{Q}^{\textsf{T}} \mathbf{Q}=\mathbf{I}_d$). Then,
$(\mathbf{Q}\boldsymbol{\alpha}_j)^{\textsf{T}} (\mathbf{Q} \boldsymbol{\beta}_i) = \boldsymbol{\alpha}_j^{\textsf{T}} \boldsymbol{\beta}_i$,
and therefore, $\textsf{Pr}(y_{ij}=1\mid\mu_j, \mathbf{Q}\boldsymbol{\alpha}_j, \mathbf{Q}\boldsymbol{\beta}_i) = \textsf{Pr}(y_{i,j}= 1 \mid \mu_j, \boldsymbol{\alpha}_j, \boldsymbol{\beta}_i)$
for all $i$ and all $j$. Such a phenomenon is also typical of latent position models for social networks \citep[e.g.,][]{Sosa-2021}.

Identifiability is not a impediment for model fitting. However, it is required to support inferences about ideal points of political agents and discrimination parameters. That is why the literature has stated necessary and sufficient conditions for the identification of spatial voting models based on paramter restrictions. For example, it is natural setting $\sigma_j=1$ because random shock scales cannot be identified separately from the scale associated with the political space \citep{yu2019circle}. On the other hand, it is common imposing restrictions on the mean and variance of the ideal points in order to make the political space invariant to translations, rotations, reflections and re-scaling \citep{jackman2004bayesian}. Specifically, $\boldsymbol{b}_{i}=\boldsymbol{0}_d$ and $\boldsymbol{B}_{i}=\mathbf{I}_d$ is fruitful to overcome simultaneously translation and scale issues. Finally, it is also convenient fixing the position of $d + 1$ legislators (also known as anchor legislators or simply anchors) with known (but distinctive!) political patterns in the political space since it allows the model to differentiate legislative tendencies (\citealp{rivers2003identification}; \citealp{clinton2004statistical}).

\subsection{Dimension of the political space}\label{sec:dim}

The choice of the dimension of the political space is a popular topic in the political science literature  (e.g., \citealp{potoski2000dimensional}; \citealp{jackman2001multidimensional}; \citealp{talbert2002setting}; \citealp{aldrich2014polarization}; \citealp{dougherty2014partisan}; \citealp{roberts2016dimensionality}; \citealp{moser2019multiple}). Research in this direction has highlighted the role of the discrimination parameters in answering questions and issues related to the number of latent features needed to model the legislators' voting behavior, the nature and meaning of the dimensions, and finally, the problems of identifiability in multidimensional models.

From a technical point of view, the dimension of the political space corresponds to the number of latent features that are necessary to characterize accurately the legislators' voting behavior. Such a choice necessarily leads to a model selection problem, in which a trade off between model fitting and model complexity needs to be faced \citep{moser2019multiple}. The available literature in this regard mostly discusses epistemological considerations (\citealp{benoit2012dimensionality}; \citealp{de2012struggle}) as well as specific quantitative mechanisms based on optimality criteria (\citealp{jackman2001multidimensional}; \citealp{lofland2017Assessing}; \citealp{moser2019multiple}). About the latter, \cite{jackman2001multidimensional} shows how discrimination parameters can be used to make a reasonable choice, and also, assess the relevance of moving to higher dimensions.

In the Latin American case, research papers determining the dimension of the political space are quite scarce. Most studies assume one or two dimensions underlying the nominal voting behavior, depending on the political context of the corresponding legislatures under analysis (e.g., \citealp{zucco2013legislative}; \citealp{zucco2011distinguishing}).  Along this road, some authors point out that the presence of religious, linguistic, or ethnic parties underlies the existence of an ideological dimension \citep{zucco2011distinguishing}. In contrast, the presence of coalitions, electoral districts, regional or provincial divisions, among others, are indicators of non-ideological dimensions (\citealp{zucco2011distinguishing}; \citealp{zucco2013legislative}). Likewise, some argue that legislative policy making is typically one-dimensional in most countries across the region \citep{rosas2005ideological}.

There is available only one work in Latin America that empirically explores the number of latent traits underlying legislative voting patterns in Argentina's parliament. \cite{jones2005party}, inspired by the work of \cite{jackman2001multidimensional}, determine the dimension of the political space through the analysis of discrimination parameters. The authors provide empirical evidence to ensure that there was only one dimension underlying the political space in the case of the roll call voting in the Argentinean House of Representatives for 1989--2003.

\subsection{Identifying pivot legislators}\label{sec:pivote}

Pivot legislators (or simply pivots, not to be confused with anchor legislators above) are those parliament members whose position in the political space is considered relevant to understand what happens within the legislature. Identifying pivot legislators is key in order to individualize the deputies' electoral behavior. Thus, research in this direction includes all sort of works focused on determining the identity and position of pivot legislators, extremists, minority legislators, among others, in order to fully characterize a given chamber. 

\cite{clinton2004statistical} state that pivot legislators are crucial to support some theories of parliamentary behavior in the North American context. This is the case because the position of these legislators in the political space can be used to characterize and predict processes of policy making (\citealp{krehbiel1998pivotal}). From a political perspective, we are not aware of whether or not pivotal theory has meaningful applications in the context of unbalanced parliaments such as the Colombian one (substantial research is required in this direction). For this reason, we limit ourselves in this work to individualize those deputies who are more likely to have either a centrist or extremist position within the parliament, as opposed to make inference on other quantities that can be obtained from ideal point estimates.

\section{Prior elicitation and computation}\label{sec:hiperYcomputo}

Now we discuss our prior choice in the case of a one-dimensional setting. Along the lines of \cite{clinton2004statistical}, we let $a_0=0$ and $A_0=25$ for the prior distribution of $\mu_{j}$ and $\alpha_{j}$, aiming to emulate roughly the behavior of a diffuse prior distribution. This choice is quite reminiscent of the hyperparameter elicitation in a stendard linear regression model when there is no need of informative or empirical alternatives, such as the unit information prior \citep{kass1996selection} or the $g$-prior \citep{zellner1986assessing}. In addition, in the same spirit of \cite{jackman2004bayesian} and \cite{lofland2017Assessing}, we let $b_i=0$ and $B_i=1$ for the prior distribution of $\beta_i$. Unlike discrimination and approval parameters, it is not necessary setting a large variance a priori for the ideal points. Actually, all what is required is a prior notion of scale for this set of parameters.

Parameter identifiability can be carried out as in \cite{rivers2003identification} and \cite{clinton2004statistical}, by selecting $d+1$ anchors (i.e., fixing the position of $d + 1$ legislators in the political space). For instance, two legislators from ideologically opposed political parties need to be anchored in our case study since we fit an one-dimensional model (see also \citealp{carroll2016unrealized}). Specifically, we delimit the positions of \textit{Jorge Enrique Robledo Castillo} (PDA member) and \textit{Roy Leonardo Barreras Montealegre} (PU member) at $-1$ and $1$, respectively. Such values are helpful to establishing a sense of scale in the political space. Thus, having fixed the location of two deputies, in a one-dimensional case like ours with $n = 91$ parliamentarians and $m = 417$ voting lists, leave us with a total of $2m + n - 2 = 923$ model parameters to estimate.

Missing data are removed from the analysis since they are not missing completely at random (MCA), so data imputation through naive sampling from the sampling distribution is not recommended (a discussion on this matter can be found in \citealp{sewell2015latent}). Although there are available imputation methods for missing data with non-random behavior (e.g., \citealp{sherina2019fully}) and even extensions of the Bayesian spatial voting model that incorporate them (e.g., \citealp{rosas2015no}), these methodologies are beyond the scope of this work (such methodologies are still not completely understood for unbalanced parliaments). However, the simulation study presented in Section \ref{sec:faltantes} shows substantial evidence that our findings are robust to different rates of missing data. 

Model fitting is carried out using MCMC algorithms (see Appendix \ref{appendix} for details) on a model based on a logit link function (see Section \ref{sec:Modelo} for details). Inferences on model parameters are based on $80,000$ samples of the posterior distribution obtained after thinning the original chains every 5 observations and a burn-in period of 24,000 iterations. In addition, before using the MCMC samples with inferential purposes, we determine first if there is any evidence of lack of convergence of any chain
to its stationary distribution. Convergence diagnostics (not shown here) show that effective sample sizes are large enough to perform adequate inductive processes.

\section{Simulation study} \label{sec:simulacion}

Here we provide an exhaustive simulation study in order to assess the robustness of a one-dimensional model to several features. To do so, each synthetic dataset is generated with $n = 91$ legislators and $m = 417$ voting lists in order to assess scenarios as close our case study as possible. Voting matrices are generated by setting the model parameters to random values. Specifically, the parameters $\alpha_j$ and $\mu_j$ are simulated from a Normal distribution with mean 0 and variance 3, whereas the ideal points $\beta_i$ are simulated from a Uniform distribution with parameters $a$ and $b$. We set $a = -3$ and $b = 3$ for balanced parliaments (two parties, each with 50\% of the deputies), and $a = -3$ and $b = 4$ for unbalanced parliaments (four parties, distributed as 75\%, group 1; 15\%, group 2; 2\%, group 3; and 8\%, group 4). Note that the unbalanced parliament emulates groups and proportions in accordance with our case study (see Section \ref{sec:apli} for details). 

The ideal points are recreated in such a way that the two groups of the balanced parliament as well as three of the unbalanced parliament (groups 1, 2, and 4) are located in a specific region of the political spectrum. This implies that members of these groups have similar ideal points. On the other hand, group 3 of the unbalanced parliament is simulated in such a way that its legislators show more heterogeneous ideal points. In particular, one of the legislators is located between groups 1 and 4, and the other between groups 2 and 4. This way of reproducing the ideal points seeks to prove that the proposed model recovers the location of the groups in the political space, regardless of low or high within-variability.

\subsection{Sensitivity analysis to anchor legislators}\label{sec:sen_leg}

In order to assess our strategy about fixing anchor legislators to $-1$ and $1$, we select such anchors from different positions across the political space (we know all the true ideal points when generating synthetic data). Thus, we consider five scenarios, namely, opposite and close to zero (Scenario 1), left--center (Scenario 2), center--right (Scenario 3), opposite and at dissimilar distances from the center (Scenario 4), and extremists (Scenario 5). We evaluate each of these possibilities for both balanced and unbalanced parliaments, using a 40\% missing data rate, and a logit link function. Particularly, Scenario 4 seems to emulate better the characteristics of the legislators that we decide to take as anchors in our case study.

\begin{table}[H]
\centering
\footnotesize
\begin{tabular}{cccccc}
\cline{3-6}
                   &                                  & \multicolumn{2}{c}{\textbf{Balanced}} & \multicolumn{2}{c}{\textbf{Unbalanced}} \\ \hline
\textbf{Scenario} & \textbf{Anchor legislators} & \textbf{DIC}       & \textbf{WAIC}       & \textbf{DIC}            & \textbf{WAIC}              \\ \hline
1                  & Opposite and close to center               & 11886.95           & 12096.09            & 8688.34                 & 8929.50           \\
2                  & Left--center               & 11798.25           & 11994.53            & 8682.17                 & 8913.65           \\
3                  & Center--right                 & 11820.27           & 12019.34            & 8573.97                 & 8802.72           \\
4                  & Different distances from the center  & 11782.73           & 11972.11            & 8563.98                 & 8775.22           \\
5                  & Extremists                      & 11809.88           & 11993.12            & 8582.86                 & 8790.65           \\ \hline
\end{tabular}
    \caption{\textit{Information criteria for contrasting different anchor legislator alternatives.}}
    \label{tab:DIC_WAIC_leg}
\end{table}

Table \ref{tab:DIC_WAIC_leg} presents the deviance information criterion (DIC; \citealp{spiegelhalter2002bayesian}) and the Watanabe--Akaike information criterion (WAIC; \citealp{watanabe2013waic}) for every scenario. Similar values are evidenced within each legislative body. Discrepancies between parliaments are to be expected since they represent two completely different chambers. In both cases, scenario 4 shows the lowest DIC and WAIC. However, these values do not show large discrepancies in comparison with the other scenarios.

The similarity among the values of the information criteria within each parliament in Table \ref{tab:DIC_WAIC_leg} strongly suggests that the model performance is not affected by the political position of the legislators used as anchor, regardless of both the number of groups and the proportions into which the parliament is divided. However, we recommend choosing as anchor legislators those parliament members known to be on opposite sides of the political spectrum (as in Scenario 4), in order to avoid selecting two legislators sharing the same ideal point, and also, ease interpretation. Even though this finding is supported by the existing literature (e.g., \citealp{lofland2017Assessing}), this is the first work that provides empirical evidence regarding the choice of anchor legislators in unbalanced parliaments considering different positions in the political space.

\begin{figure}[!t]
\centering
    \subfigure[$\hat{\beta}_i$ vs. $\beta_i$, balanced parliament.]{
    \centering
        \includegraphics[scale=0.31]{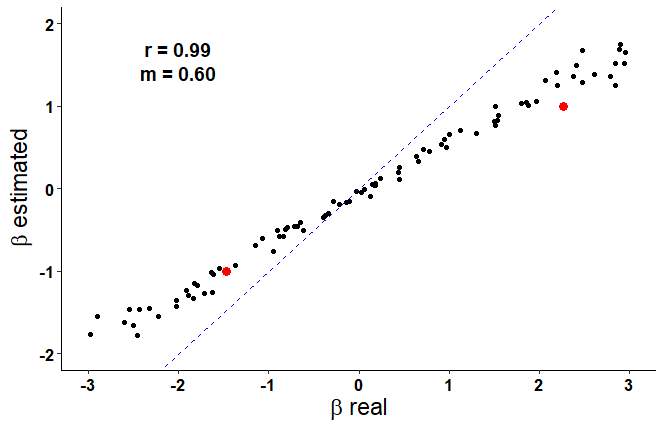}
        \label{fig:sim_leg1}}
    \hfill
    \subfigure[$\hat{\beta}_i$ vs. $\beta_i$, unbalanced parliament.]{
    \centering
        \includegraphics[scale=0.31]{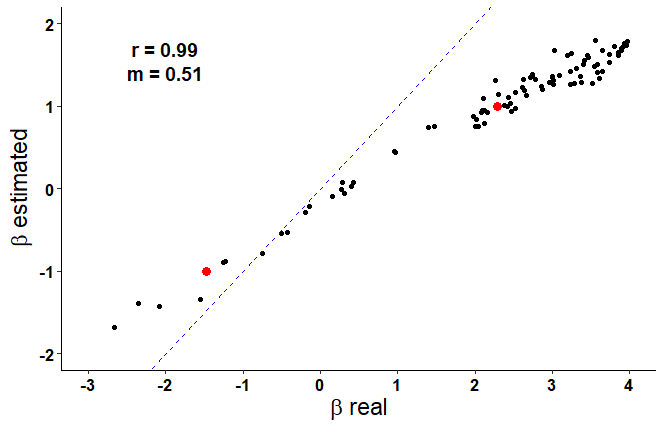}
        \label{fig:sim_leg2}}
    \caption{\textit{Scatter plots of $\beta_i$ vs. $\hat \beta_i$ under Scenario 4 (different distances from the center). Red points indicate the location of legislators used as anchors, whereas blue lines represent the line of reference $\beta = \hat \beta$. Finally, $r$ is the Pearson's correlation coefficient, and $m$ is the slope associated with the corresponding regression line.}}
    \label{fig:sim_leg}
\end{figure}

Figure \ref{fig:sim_leg} makes it clear that the deputies’ ideal points are not accurately recovered because the model is not identifiable, and also, parameter estimation is carried out on a different scale (see Sections \ref{sec:identi} and \ref{sec:hiperYcomputo} for details). Also, we see that there is a greater scaling factor effect on extreme ideal points, as opposed to those located around the center. Extreme values of discrimination and approval parameters (not shown here) are also more variable because of the scale effect.
Finally, note that scales behave proportionally with a constant rate of change that can be quantified by the slope of the corresponding regression line.

\begin{figure}[!b]
\centering
    \subfigure[Ideal points, balanced parliament.]{
    \centering
        \includegraphics[scale=0.31]{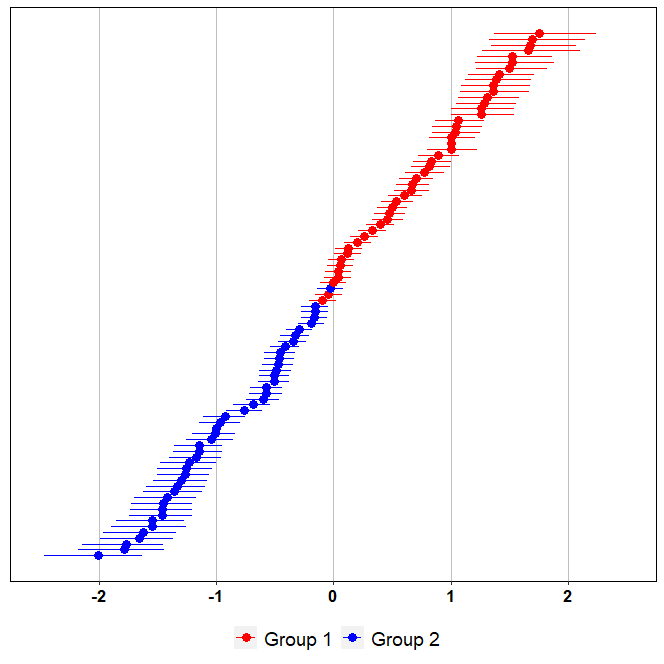}
        \label{fig:ideal1}}
    \hfill
    \subfigure[Ideal points, unbalanced parliament.]{
    \centering
        \includegraphics[scale=0.31]{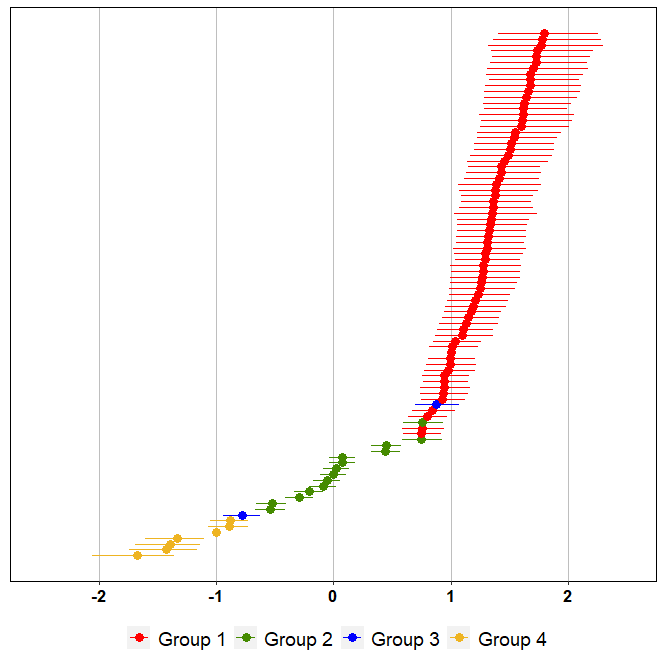}
        \label{fig:ideal2}}
    \caption{\textit{Estimates under the election of opposite anchor legislators and at different distances from the center (scenario 4). In each case, filled circles represent point estimates, and horizontal lines the corresponding 95\% credible interval based on percentiles.}}
    \label{fig:ideal}
\end{figure}

Despite the change the scaling issues, by anchoring two legislators at $-1$ and $1$, we are able to recover groups as they were generated no matter what the scenario is. Such a behavior is particularly evident in Figure \ref{fig:ideal} where we show our findings for Scenario 4. Specifically, we see that for both parliaments, ideal point estimates between $- 1$ and $1$ reveal smaller credibility intervals. As legislators move away from the center towards the extremes of the political spectrum, the uncertainty about the ideal point estimates grows. This implies that the rescaling effect is mostly evident through the ideal points' margin of error. Also, legislators from different groups, whose ideal points are similar, are the most likely to be mixed with each other. The exchanges that are evident among members of different groups (e.g., around zero in Figure \ref{fig:ideal}a) derive from the inherent uncertainty in the estimation of the parameters.

\subsection{Sensitivity analysis to link functions} \label{sec:eleccionEnlace}

In order to investigate the robustness of the analysis to the choice of link functions (see Section \ref{sec:Modelo} for details). To do so, we consider an unbalanced parliament with a 40\% missing data rate each time, and generate voting data using a logit link. Then, we fit the model anchoring to $-1$ and $1$ two opposite legislators at different distances from the center, and also, using both a logit link (as in Scenario 4 above) as well as a probit link (Scenario 6). Actually, all the different combinations of data generating process/model fitting in terms of link functions, namely, logit--logit, logit--probit, probit--logit, probit--probit, yield to analogous results. 

The information criteria in Scenario 4 (DIC $=8563.98$, WAIC $=8775.22$) and Scenario 6 (DIC $=8420.61$, WAIC $=8670.81$) reflect a similar results in predictive terms. Although the probit model exhibits a lower information criteria than the logit model, such a difference does not make clear an advantage of one model over the other. When comparing the true value of the ideal points with their corresponding estimates, we observe a similar behavior in either case, since both links generate parameter estimates at a lower scale (which is mostly and effect of our choice to make the model identifiable).
Finally, note that our findings are consistent with that stated by other authors in terms of the choice of link functions for univariate binary response models (e.g., \citealp{hahn2005probit}), particularly in the case of balanced parliaments (e.g., \citealp{lofland2017Assessing}).

\subsection{Sensitivity Analysis to the missing data rate} \label{sec:faltantes}

In order to evaluate the performance of the model at different missing data rates, we contrasted three scenarios, namely, with a missing rate of 10\% (Scenario 7), 40\% (as in Scenario 4 above), and 60\% (Scenario 8). For this comparison, we use an unbalanced parliament taking as anchors legislators of opposite anchor and at different distances from the center, and using the logit function as a link.

\begin{figure}[!t]
\centering
    \subfigure[Scenario 7]{
    \centering
        \includegraphics[scale=0.23]{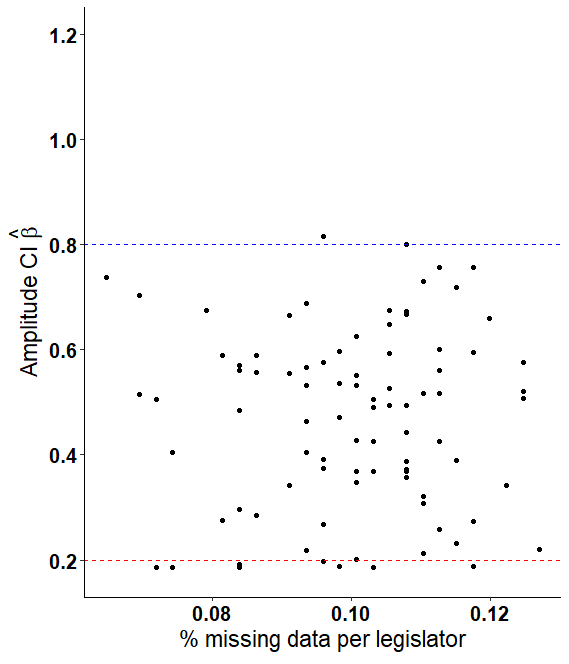}
        \label{fig:beta1}}
    \hfill
    \subfigure[Scenario 4]{
    \centering
        \includegraphics[scale=0.23]{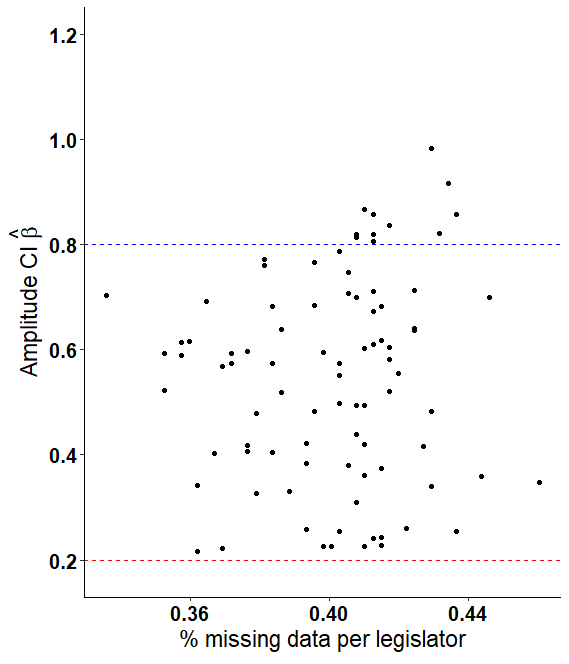}
        \label{fig:beta4}}
    \hfill    
    \subfigure[Scenario 8]{
    \centering
        \includegraphics[scale=0.23]{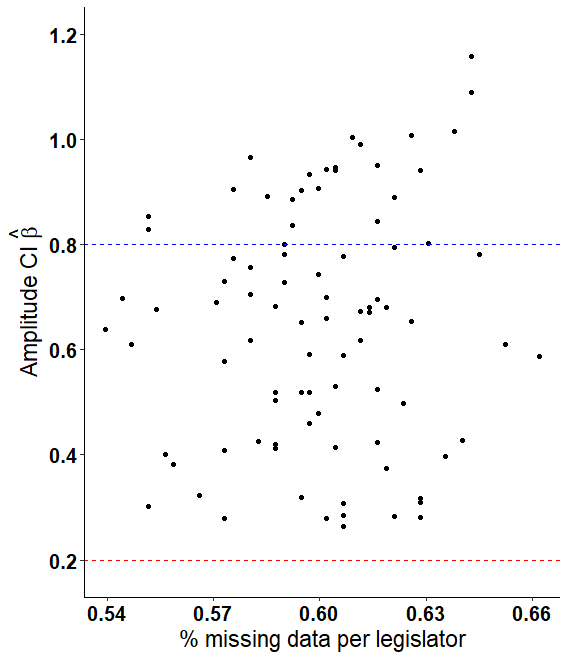}
        \label{fig:beta6}}
    \caption{\textit{Scatterplots for ideal points' 95\% credibility interval amplitude vs. missing data rate per legislator, corresponding to scenarios 7 (10\% missing data rate), 4 (40\% missing data rate), and 8 (60\% missing data rate). Lines are arbitrary and drawn to ease visualization.}}
    \label{fig:amplitud}
\end{figure}

Figure \ref{fig:amplitud} shows that an increase in the percentage of missing data implies an increase in the width of the credibility intervals of the ideal points (the same pattern is revealed with the discrimination and approval parameters, when the width is contrasted with the percentage of missing data per voting list). This observation allows us to conclude that the higher the rate of missing data, the greater the uncertainty in the estimates of all the parameters of the model. Furthermore, the missing data impact seems to be larger for those estimates that show values at the extremes of the political spectrum. Interestingly, we see that different missing data rates do not generate erroneous conclusions in terms of inference about the model parameters. In particular, it does not distort the clustering of the ideal points. In fact, they continue to show the pattern in Figure \ref{fig:ideal}b.

\subsection{Sensitivity analysis to prior distributions of ideal points}

Finally, we investigate the robustness of the analysis to the choice of prior distribution and hyperparameters for the ideal points. Such a task is very important because later analyses focus on these parameters. The comparison of the scenarios proposed below is done for an unbalanced parliament with a 40\% missing data rate taking as anchors legislators of opposite anchor and at different distances from the center, and using the logit function as a link.

Three new scenarios are contrasted, namely, prior distribution without hierarchies (as in Scenario 4 above), i.e., $b_i=0$ y $B_{i}=1$; hierarchical prior with two stages for the variance of ideal points (Scenario 9), i.e., $b_i=0$ y $B_{i} \sim \textsf{GI}(c,d)$, with $c = 3$ and $d = 2$; hierarchical prior with two stages on the mean and variance (Scenario 10), i.e., $b_i \sim \textsf{N}(a,b)$ y $B_{i} \sim \textsf{GI}(c,d)$, with $a=0$ and $b=25$. The choices described lead to diffuse priors on the ideal points, since if $c = 3$ and $d = 2$, then the coefficient of variation of the Inverse Gamma distribution is $1$ (see also the discussion in this regard provided in Section \ref{sec:hiperYcomputo}). The hyperparameters of the discrimination and approval parameters are $a_0 =0$ and $A_0 = 25$.

The information criteria (Scenario 4, DIC $=8563.98$, WAIC $=8775.22$; Scenario 9, DIC $=8555.55$, WAIC $=8774.05$; Scenario 10, DIC $=8553.48$, WAIC $=8771.59$) show that Scenario 10 presents a better predictive performance. However, note once again that the corresponding improvement is not substantial. Therefore, such hierarchization can be dispensed with in order to preserve parsimony, considering that under the three scenarios, the groups are recovered on a similar scale, and in all cases, the pattern of Figure \ref{fig:ideal}b is repeated.

\section{Senate of the Republic of Colombia 2010--2014} \label{sec:apli}

The Sixth Congress of the Republic of Colombia 2010-2014 is characterized mainly by two aspects: (i) The leadership of the National Unity, a coalition of political parties formed to support the first government of Juan Manuel Santos, and (ii) a substantial number of motions proposed for deliberation. Several parties take part in this particular Senate. \textit{Partido Social de unidad Nacional} (PU), \textit{Conservador Colombiano} (CC), \textit{Liberal Colombiano} (LC) and \textit{Cambio Radical} (CR) are those collectivities that make up the government coalition, whereas \textit{Partido Polo Alternativo Democrático} (PDA) is the only one in open opposition. Furthermore, \textit{Partido Integración Nacional} (PIN), \textit{Partido Alianza Verde} (PAV), and \textit{MIRA} are independent political groups. Finally, \textit{Alianza Social Indígena} (ASI) along with \textit{Autoridades Indígenas de Colombia} (AICO) are political parties representing minorities. The extensive legislative activity, the partisan configuration, and the importance of constitutional reforms and bills in the context of the peace process (e.g., \citealp{deVisibleOsorio2014}) make of this a four-year period worth of analysis in terms of the legislative behavior of deputies as well as the partisan and coalition patterns to take legislative decisions.

For evaluating the political preferences underlying the voting behavior of deputies, we use the plenary voting records available for the period 2010--2014. The choice of plenary votes obeys the intention of providing inferences based on the observed voting behavior of deputies when they act on the same voting lists. This decision is also supported by the fact that plenary votes are the most relevant activity of the parliament for different groups. In particular, they are votes that capture the attention of political groups, the media, and voters, because of the impact of their results on state policy \citep{aleman2008policy}.

Voting reports on bills can be obtained via the Visible Congress Web at \url{https://congresovisible.uniandes.edu.co/}. The dataset provides information on the bill, legislators, and voting decisions. Legislators taking part in less than 95\% of Senate plenary votes for 2010--2014 are excluded. They correspond to 8\% of the total number of deputies (102 permanent deputies and 8 permanent replacements). Their low participation results from different causes: death, resignation, disciplinary sanctions, or permanent replacement. Eliminating these deputies does not imply eliminating any political group.

\subsection{Ideal points patterns}

\begin{figure}[!t]
\centering
    \includegraphics[scale=0.64]{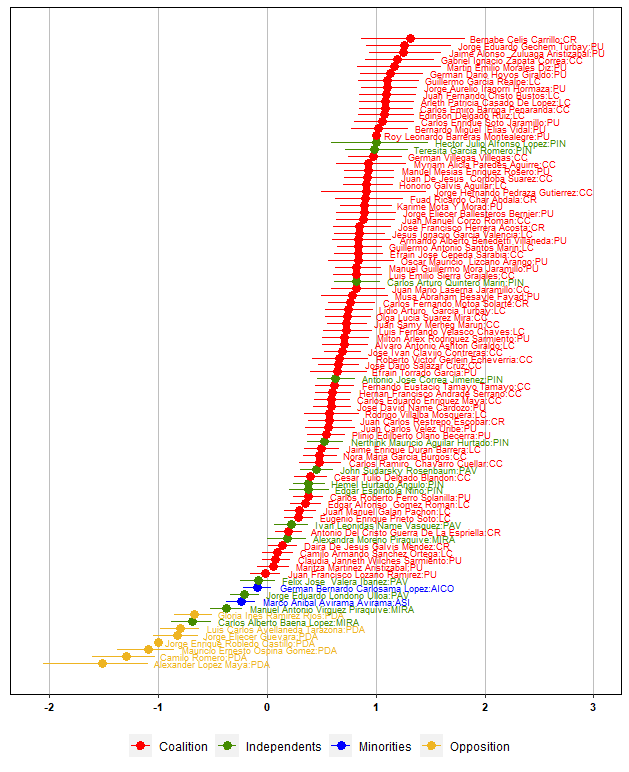}
    \caption{\textit{Senate ideal points estimates under the election of presumed opposing legislators and at different distances from the center as anchors. Points represent the posterior mean, and horizontal lines 95\% symmetric credibility interval based on percentiles.}}
    \label{fig:id_senado}
\end{figure}

We fit a Bayesian spatial voting model to these data as discussed in Section \ref{sec:hiperYcomputo}. Figure \ref{fig:id_senado} shows the ideal points estimates of the Senate deputies that take part of the analysis. The estimates exhibit a particular behavior according to the political group in which the deputies are enrolled. For example, members of the opposition reveal a location on the left side of the political spectrum. This position is opposite to that of the members of the ruling coalition, whose estimated ideal points, for the most part, are greater than zero. Minorities reveal a center--left location (between $- 0.5$ and $0$), whereas independent parties are observed to be more dispersed across the political space. The opposition and the coalition are the political groups with the least within-variability (the coefficient of variation of the ideal points of their members is 29\% and 42\%, respectively). Minorities and independents are the most heterogeneous groups (the coefficient of variation of the ideal points of their members is 62\% and 167\%, respectively).

Deputies with ideal points located towards the extreme ends of the political spectrum and with high percentages of abstention and non--attendance, are those who reveal wider credibility intervals, thus, greater uncertainty in the estimation of their positions. Such is the case of senators Alexander López Maya of the PDA, Jorge Hernando Pedraza Gutierrez of the CC, and Hector Julio Alfonso López of the PIN, who present high abstention percentages and exhibit ideal points distant from the center of the political space.

\begin{figure}[!t]
\centering
    \includegraphics[scale=0.65]{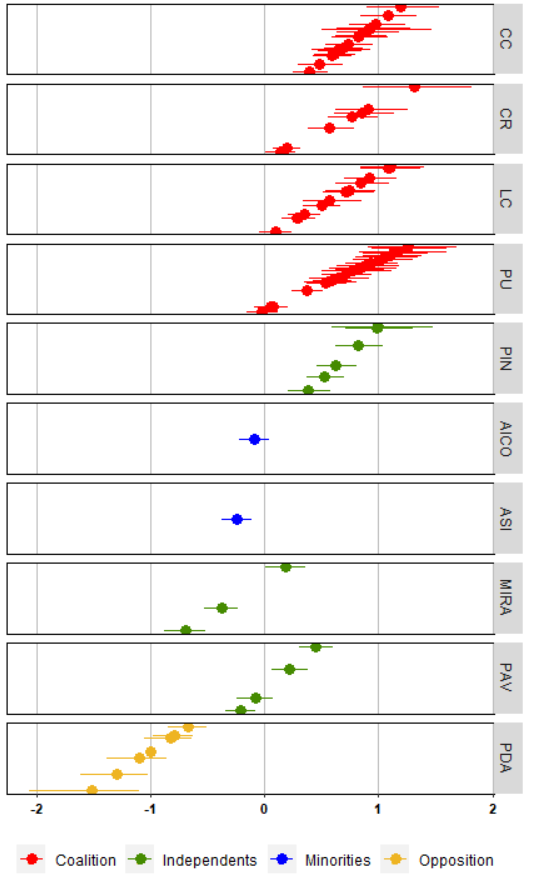}
    \caption{\textit{Classification of the Senate's ideal points estimates by political party. Points represent the posterior mean, and horizontal lines 95\% credibility interval based on percentiles.}}
    \label{fig:partido_senado}
\end{figure}

The credibility intervals with the smallest amplitude are recorded for legislators whose ideal points are in $[- 0.5, 0.5]$, which reflects less uncertainty in the corresponding positions of centrist parliamentarians. Legislator Jorge Eliecer Guevara of the PDA, who changed parties during the course of his tenure (he ends his four--year term in the PAV), reveals a position that is consistent with the political group in which he begins his legislative work. Also, 83 out of 89 ideal points are significantly different from zero (i.e., the corresponding credibility interval does not contain zero). Legislators German Bernardo Carlosama López of AICO, Feliz Jose Varela Ibañez of PAV, Camilo Armando Sanchez Ortega of LC, Juan Francisco Lozano Ramirez, Maritza Martínez Aristizabal, and Claudia Janneth Wilches Sarmiento of PU have ideal points indistinguishable from zero.

\begin{figure}[!t]
\centering
    \subfigure[$\alpha_j$ discriminating in one dimension.]{
    \centering
        \includegraphics[scale=0.32]{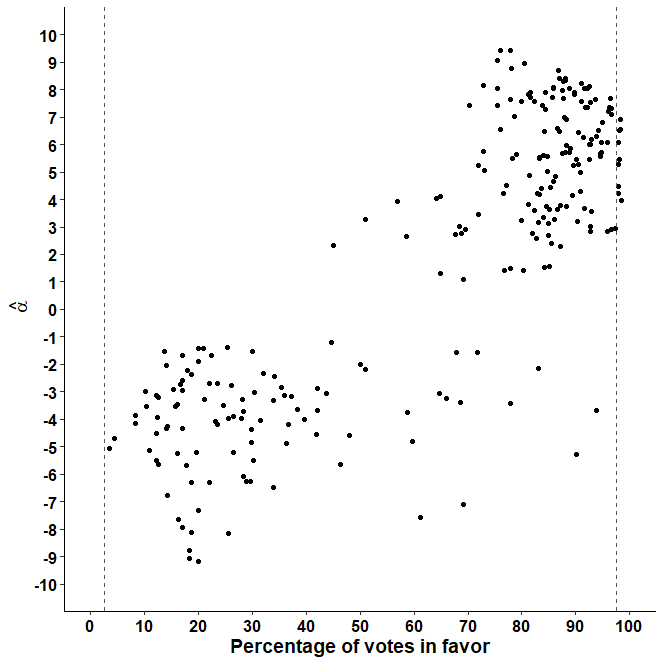}
        \label{fig:discri}}
    \hfill
    \subfigure[$\alpha_j$ not discriminating in one dimension.]{
    \centering
        \includegraphics[scale=0.32]{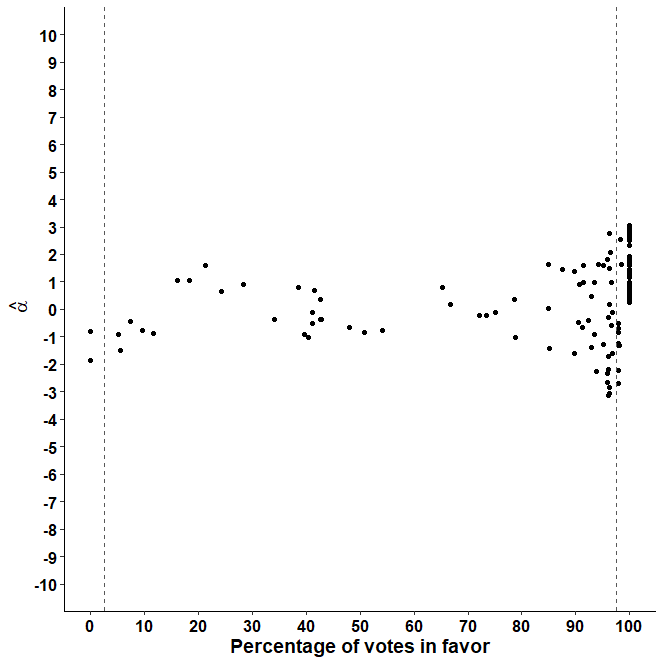}
        \label{fig:no_discri}}
    \caption{\textit{Scatterplots of discrimination parameter estimates vs. percentage of votes in favor the corresponding voting list. Vertical lines in gray mark the unbalanced elections (voting lists with a percentage of votes in favor less than or equal to 2.5\%, or greater than or equal to 97.5\%).}}
    \label{fig:discriminacion}
\end{figure}

Senators with a higher probability of being at the extreme ends of the political spectrum are those belonging to the opposition or a coalition party. From the PDA legislators, those with a high probability of having an ideal point lower than $- 1$ are: Alexander López Maya (99\%), Camilo Romero Galeano (98\%), and Mauricio Ernesto Ospina Gomez (75\%). From the coalition legislators, those who have a high probability of having an ideal point greater than $1$ are: Jaime Alonso Zuluaga Aristizabal (93\%) and Jorge Eduardo Gechem Turbay (92\%) of the PU, Bernabe Celis Carrillo of CR (90\%), and Gabriel Ignacio Zapata Correa of the CC (89\%). Legislators from the LC register a probability of 76\% that being at the upper end of the political spectrum. Some members of the minorities and independent parties present a high probability of being at the center of the spectrum. In particular, the deputies with a high probability of having an ideal point between $- 0.2$ and $0.2$ (i.e., having locations close to zero) are: Germán Bernardo Carlosama López (95\%) of AICO and Feliz Jose Valera Ibañez (92\%) of PAV. All PIN members have a probability of at least 98\% of having an ideal point higher than $0.2$. The latter indicates that legislators affiliated to the PIN have a high tendency towards the government coalition, since they are distant from the opposition and centrist legislators. 

Figure \ref{fig:partido_senado} shows the behavior of the ideal points by political party. A similar pattern is observed in the members of the collectivities that make up the coalition (CC, LC, CR, PU) and PIN. The latter does not declare itself to be in favor or against the government, but its political trajectory has shown an affinity with parties such as the PU \citep{pin}. Therefore, it is to be expected that the ideal points of its members are aligned with those of the government parties. The PDA, the only party in opposition, reveals ideal points at the opposite extreme of the government parties and PIN. The parties with fewer seats in the Senate (AICO, ASI, MIRA, and PAV) show ideal points around zero. The minorities show center--left tendencies (ideal points between $-0.5$ and $0$), and MIRA as well as PAV exhibit senators with center--left and center--right locations (ideal points between $0$ and $0.5$).

\subsection{Dimension of the political space via discrimination parameters}

In terms of the discrimination parameters estimates, we see that 251 of the 417 differ significantly from zero. This implies that, under the proposed unidimensional model, there are 60.2\% of the motions that can discriminate among legislators across the political spectrum. 
Meanwhile, no relevant topic pattern is identified regarding the dimension of the political space. Any other way, we could determine that 81.7\% of the motions that discriminate against a dimension are those about governmental initiatives, a finding that is consistent with the proposed latent trait (opposition--non-opposition).

Figure \ref{fig:discriminacion} shows the joint behavior of the percentage of votes to support the voting list $j$ and the respective point estimate of the discrimination parameter $\alpha_j$. By equating the behaviors of the $\alpha_j$ that are distinguished from zero (Figure \ref{fig:discriminacion}a) with those that do not segregate in the proposed dimension (Figure \ref{fig:discriminacion}b), we can observe that voting lists with 0\% and 100\% (unanimous votes) do not distinguish between legislators. Therefore, they are not relevant to recover information about the political preferences of the deputies. Futhermore, 89 out of 166 non-segregating motions correspond to votes of this type. The remaining 77 possibly contain information about an additional latent trait underlying the roll call of parliamentarians.

Voting lists that discriminate among deputies having a percentage of votes in favor less than 45\% reveal negative discrimination parameters (Figure \ref{fig:discriminacion}a). This implies that legislators with ideal points lower than zero increase their probability of positive vote, while legislators with ideal points higher than zero decrease this probability. An inverse pattern is highlighted for voting lists with a percentage of vote in favor higher than 70\% (Figure \ref{fig:discriminacion}a). Voting lists with a positive vote percentage between 40\% and 70\% are more likely to be dispersed around zero.

Finally, we obtain the posterior predictive distribution for a set of test statistics (see \citealp{gelman2013bayesian} for details). All subsequent predictive $p$-values provide reasonable evidence supporting a good fit, since they do not correspond to extreme values (less than 0.05 or greater than 0.95). Furthermore, all distributions (not shown here) show that the observed test statistics values falls within the 95\% symmetric credibility interval. In this sense, such a predictive tests gives conclusive evidence that the proposed one-dimensional model reveals a reasonable fit to the roll call data about the Senate of the Republic of Colombia 2010--2014.

\section{Discussion} \label{sec:con}

Our simulations study provides strong evidence about the conditions under which the results obtained after fitting a one-dimensional Bayesian ideal point  model to nominal voting data of an unbalanced parliament are invariant. In particular, we conclude that the choice of anchor legislators can be arbitrary and that there are no substantial differences between logit and probit links working in a univariate context. It is up to later studies to analyze the sensitivity of these binding functions when working in higher dimensions and even under non--parametric conditions \citep{newton1996bayesian}.

The sensitivity analysis to the prior distributions of the ideal points indicates that the results are consistent with hyperparameters with and without extra hierarchies. In future studies, it is of interest to inquire about the sensitivity of the model when other parametric or non-parametric families are considered for the hyperparameters of the model, including those associated with the discrimination and approval parameters. On the other hand, it is evident that dispensing with the missing data does not generate erroneous conclusions in terms of the inference of the model parameters. However, a higher rate of missing data generates a greater uncertainty in terms of estimation. The 2010--2014 Senate abstentions and non--attendance rate is high (approximately 40\%), which prompts further studies that inquire about the mechanism that generates the lost data (e.g., \citealp{rosas2015no}).

This paper is the first in its class to apply the Bayesian ideal point estimator in the Colombian context. The results are a substantial contribution to the topic regarding the dimension of political space and identification of pivot legislators. The proposed methodology allows us to identify those legislators who have a higher probability of being at the extremes or in the center of the political spectrum, but does not determine the order in which parliamentarians are placed (e.g., \citealp{clinton2004statistical}). Legislative analysis at the individual level allows us to personalize patterns of electoral behavior and include legislators from political parties with a low number of seats in parliament, such as AICO, ASI, PAV, and MIRA, which are usually excluded in quantitative research that only extrapolates at the group level.

By means of the pattern that reveals the estimated ideal points, we can recognize a non-ideological latent trait (opposition--non-opposition) underlying the voting behavior of Senate deputies. This does not rule out the possibility of extending the one--dimensional standard Bayesian ideal point estimator to assess other latent factors in the voting behavior of parliamentarians. The estimator in its canonical version operates under the assumption of sincere voting (i.e., the legislator only votes according to his or her ideal point) and therefore, it does not allow to measure ideological, partisan, or coalition effects in deputy voting (e.g., \citealp{tsai2020influence}).

Finally, the one-dimensional model exhibits a reasonable fit to the nominal voting data of the Senate of the Republic of Colombia 2010--2014. However, we identified some voting lists that do not discriminate among legislators on the continuum of recovered policies. This finding raises the possibility of evaluating a model in higher dimensions to examine whether it is justified to increase the complexity of the model in order to get a better fit to the roll call voting data of this legislative chamber. It also leaves an open possibility of evaluating the static and collective character of the political space dimension in the Colombian context (e.g., \citealp{moser2019multiple}). 

\bibliography{references}

\begin{thebibliography}{}

\bibitem[Albert and Chib, 1993]{albert1993bayesian}
Albert, J.~H. and Chib, S. (1993).
\newblock Bayesian analysis of binary and polychotomous response data.
\newblock {\em Journal of the American statistical Association},
  88(422):669--679.

\bibitem[Aldrich et~al., 2014]{aldrich2014polarization}
Aldrich, J.~H., Montgomery, J.~M., and Sparks, D.~B. (2014).
\newblock Polarization and ideology: Partisan sources of low dimensionality in
  scaled roll call analyses.
\newblock {\em Political Analysis}, pages 435--456.

\bibitem[Alem{\'a}n, 2008]{aleman2008policy}
Alem{\'a}n, E. (2008).
\newblock Policy positions in the chilean senate: An analysis of coauthorship
  and roll call data.
\newblock {\em Brazilian Political Science Review (Online)}, 3(SE):0--0.

\bibitem[Alem{\'a}n and Pach{\'o}n, 2008]{aleman2008comisiones}
Alem{\'a}n, E. and Pach{\'o}n, M. (2008).
\newblock Las comisiones de conciliaci{\'o}n en los procesos legislativos de
  chile y colombia.
\newblock {\em Pol{\'\i}tica y gobierno}, 15(1):03--34.

\bibitem[Archer and Shugart, 1997]{archer1997unr}
Archer, R.~P. and Shugart, M.~S. (1997).
\newblock The unrealized potential of presidential dominance in {C}olombia.
\newblock {\em Presidentialism and democracy in Latin America}, pages 110--160.

\bibitem[Arrow, 1990]{arrow1990advances}
Arrow, K. (1990).
\newblock {\em Advances in the spatial theory of voting}.
\newblock Cambridge University Press.

\bibitem[Benoit and Laver, 2012]{benoit2012dimensionality}
Benoit, K. and Laver, M. (2012).
\newblock The dimensionality of political space: Epistemological and
  methodological considerations.
\newblock {\em European Union Politics}, 13(2):194--218.

\bibitem[C{\'a}rdenas et~al., 2008]{cardenas2008political}
C{\'a}rdenas, M., Junguito, R., and Pach{\'o}n, M. (2008).
\newblock Political institutions and policy outcomes in colombia: The effects
  of the 1991 constitution.
\newblock {\em Policymaking in Latin America: how politics shapes policies},
  pages 199--242.

\bibitem[Carroll and Pach{\'o}n, 2016]{carroll2016unrealized}
Carroll, R. and Pach{\'o}n, M. (2016).
\newblock The unrealized potential of presidential coalitions in {C}olombia.
\newblock {\em Legislative Institutions and Lawmaking in Latin America}, pages
  122--147.

\bibitem[Clinton et~al., 2004]{clinton2004statistical}
Clinton, J., Jackman, S., and Rivers, D. (2004).
\newblock The statistical analysis of roll call data.
\newblock {\em American Political Science Review}, pages 355--370.

\bibitem[Colprensa, 2013]{pin}
Colprensa (2013).
\newblock El {PIN} cambi\'o el nombre de su partido a opci\'on {C}iudadana.
\newblock {\em El pa\'is}.

\bibitem[Croux et~al., 2004]{croux2004robust}
Croux, C., Dhaene, G., and Hoorelbeke, D. (2004).
\newblock Robust standard errors for robust estimators.
\newblock {\em CES-Discussion paper series (DPS) 03.16}, pages 1--20.

\bibitem[De~Vries and Marks, 2012]{de2012struggle}
De~Vries, C.~E. and Marks, G. (2012).
\newblock The struggle over dimensionality: A note on theory and empirics.
\newblock {\em European Union Politics}, 13(2):185--193.

\bibitem[Dougherty et~al., 2014]{dougherty2014partisan}
Dougherty, K.~L., Lynch, M.~S., and Madonna, A.~J. (2014).
\newblock Partisan agenda control and the dimensionality of congress.
\newblock {\em American Politics Research}, 42(4):600--627.

\bibitem[Downs, 1957]{downs1957economic}
Downs, A. (1957).
\newblock An economic theory of political action in a democracy.
\newblock {\em Journal of political economy}, 65(2):135--150.

\bibitem[Gamerman and Lopes, 2006]{gamerman2006markov}
Gamerman, D. and Lopes, H. (2006).
\newblock {\em Markov chain Monte Carlo: stochastic simulation for Bayesian
  inference}.
\newblock CRC Press.

\bibitem[Gelman et~al., 2014]{gelman2013bayesian}
Gelman, A., Carlin, J.~B., Stern, H.~S., Dunson, D.~B., Vehtari, A., and Rubin,
  D.~B. (2014).
\newblock {\em Bayesian data analysis}.
\newblock CRC press.

\bibitem[Hahn and Soyer, 2005]{hahn2005probit}
Hahn, E.~D. and Soyer, R. (2005).
\newblock Probit and logit models: Differences in the multivariate realm.
\newblock {\em The Journal of the Royal Statistical Society, Series B}, pages
  1--12.

\bibitem[Hare et~al., 2015]{hare2015using}
Hare, C., Armstrong, D.~A., Bakker, R., Carroll, R., and Poole, K.~T. (2015).
\newblock Using bayesian aldrich-mckelvey scaling to study citizens'
  ideological preferences and perceptions.
\newblock {\em American Journal of Political Science}, 59(3):759--774.

\bibitem[Hoskin, 1975]{hoskin1975dimensions}
Hoskin, G. (1975).
\newblock Dimensions of conflict in the colombian national legislature.
\newblock {\em Legislative Systems in Developing Countries}, pages 143--178.

\bibitem[Hoskin, 1979]{hoskin1979belief}
Hoskin, G. (1979).
\newblock Belief systems of colombian political party activists.
\newblock {\em Journal of Interamerican Studies and World Affairs},
  21(4):481--504.

\bibitem[Hoskin et~al., 1976]{hoskin1976legislative}
Hoskin, G., Kline, H.~F., and Buitrago, F.~L. (1976).
\newblock {\em Legislative Behavior in Colombia}.
\newblock Council on International Studies, State University of New York at
  Buffalo.

\bibitem[Hoskin and Swanson, 1973]{hoskin1973inter}
Hoskin, G. and Swanson, G. (1973).
\newblock Inter-party competition in colombia: a return to la violencia?
\newblock {\em American Journal of Political Science}, pages 316--350.

\bibitem[Jackman, 2001]{jackman2001multidimensional}
Jackman, S. (2001).
\newblock Multidimensional analysis of roll call data via bayesian simulation:
  Identification, estimation, inference, and model checking.
\newblock {\em Political Analysis}, 9(3):227--241.

\bibitem[Jackman, 2004]{jackman2004bayesian}
Jackman, S. (2004).
\newblock Bayesian analysis for political research.
\newblock {\em Annu. Rev. Polit. Sci.}, 7:483--505.

\bibitem[Jones and Hwang, 2005]{jones2005party}
Jones, M.~P. and Hwang, W. (2005).
\newblock Party government in presidential democracies: Extending cartel theory
  beyond the us congress.
\newblock {\em American Journal of Political Science}, 49(2):267--282.

\bibitem[Kass and Wasserman, 1996]{kass1996selection}
Kass, R.~E. and Wasserman, L. (1996).
\newblock The selection of prior distributions by formal rules.
\newblock {\em Journal of the American statistical Association},
  91(435):1343--1370.

\bibitem[Kline, 1974]{kline1974interest}
Kline, H.~F. (1974).
\newblock Interest groups in the colombian congress: Group behavior in a
  centralized, patrimonial political system.
\newblock {\em Journal of Interamerican Studies and World Affairs},
  16(3):274--300.

\bibitem[Kline, 1977]{kline1977committee}
Kline, H.~F. (1977).
\newblock Committee membership turnover in the colombian national congress,
  1958-1974.
\newblock {\em Legislative Studies Quarterly}, pages 29--43.

\bibitem[Krehbiel, 1988]{krehbiel1988spatial}
Krehbiel, K. (1988).
\newblock Spatial models of legislative choice.
\newblock {\em Legislative Studies Quarterly}, pages 259--319.

\bibitem[Krehbiel, 1998]{krehbiel1998pivotal}
Krehbiel, K. (1998).
\newblock {\em Pivotal politics: A theory of US lawmaking}.
\newblock University of Chicago Press.

\bibitem[Lofland et~al., 2017]{lofland2017Assessing}
Lofland, C.~L., Rodr{\'\i}guez, A., Moser, S., et~al. (2017).
\newblock Assessing differences in legislators’ revealed preferences: A case
  study on the 107th us senate.
\newblock {\em The Annals of Applied Statistics}, 11(1):456--479.

\bibitem[Moser et~al., 2021]{moser2019multiple}
Moser, S., Rodr{\'\i}guez, A., and Lofland, C.~L. (2021).
\newblock Multiple ideal points: Revealed preferences in different domains.
\newblock {\em Political Analysis}, 29(2):139--166.

\bibitem[Newton et~al., 1996]{newton1996bayesian}
Newton, M.~A., Czado, C., and Chappell, R. (1996).
\newblock Bayesian inference for semiparametric binary regression.
\newblock {\em Journal of the American Statistical Association},
  91(433):142--153.

\bibitem[Osorio, 2014]{deVisibleOsorio2014}
Osorio, C. (2014).
\newblock Unidad nacional: cr{\'i}tica, pero muy {\'u}til.
\newblock {\em Congreso Visible}.

\bibitem[Pachón and Muñoz, 2020]{pachon2020}
Pachón, M. and Muñoz, M. (2020).
\newblock {\em Policy analysis and the legislature in Colombia}, pages 81--98.
\newblock Bristol University Press, first edition.

\bibitem[Pach{\'o}n, 2011]{pachon2011que}
Pach{\'o}n, M. (2011).
\newblock ?` qu{\'e} tanta pol{\'\i}tica nacional discute un congreso? una
  comparaci{\'o}n de las agendas de las plenarias y comisiones posterior a la
  constituci{\'o}n de 1991.
\newblock {\em Revista latinoamericana de pol{\'i}tica comparada}, 4:75--98.

\bibitem[Pach{\'o}n and Johnson, 2016]{pachon2016s}
Pach{\'o}n, M. and Johnson, G.~B. (2016).
\newblock When's the party (or coalition)? agenda-setting in a highly
  fragmented, decentralized legislature.
\newblock {\em Journal of Politics in Latin America}, 8(2):71--100.

\bibitem[Poole, 2000]{poole2000nonparametric}
Poole, K.~T. (2000).
\newblock Nonparametric unfolding of binary choice data.
\newblock {\em Political Analysis}, 8(3):211--237.

\bibitem[Potoski and Talbert, 2000]{potoski2000dimensional}
Potoski, M. and Talbert, J. (2000).
\newblock The dimensional structure of policy outputs: Distributive policy and
  roll call voting.
\newblock {\em Political Research Quarterly}, 53(4):695--710.

\bibitem[Rivers, 2003]{rivers2003identification}
Rivers, D. (2003).
\newblock Identification of multidimensional item-response models.
\newblock {\em Typescript. Department of Political Science, Stanford
  University}.

\bibitem[Roberts et~al., 2016]{roberts2016dimensionality}
Roberts, J.~M., Smith, S.~S., and Haptonstahl, S.~R. (2016).
\newblock The dimensionality of congressional voting reconsidered.
\newblock {\em American Politics Research}, 44(5):794--815.

\bibitem[Rosas, 2005]{rosas2005ideological}
Rosas, G. (2005).
\newblock The ideological organization of latin american legislative parties:
  An empirical analysis of elite policy preferences.
\newblock {\em Comparative Political Studies}, 38(7):824--849.

\bibitem[Rosas et~al., 2015]{rosas2015no}
Rosas, G., Shomer, Y., and Haptonstahl, S.~R. (2015).
\newblock No news is news: Nonignorable nonresponse in roll-call data analysis.
\newblock {\em American Journal of Political Science}, 59(2):511--528.

\bibitem[Sewell and Chen, 2015]{sewell2015latent}
Sewell, D.~K. and Chen, Y. (2015).
\newblock Latent space models for dynamic networks.
\newblock {\em Journal of the American Statistical Association},
  110(512):1646--1657.

\bibitem[Sherina et~al., 2019]{sherina2019fully}
Sherina, V., McCall, M.~N., and Love, T.~M. (2019).
\newblock Fully bayesian imputation model for non-random missing data in qpcr.
\newblock {\em arXiv preprint arXiv:1910.13936}.

\bibitem[Sosa and Buitrago, 2021]{Sosa-2021}
Sosa, J. and Buitrago, L. (2021).
\newblock A review of latent space models for social networks.
\newblock {\em Revista Colombiana de Estad{\'\i}stica}, 44(1):171--200.

\bibitem[Spiegelhalter et~al., 2002]{spiegelhalter2002bayesian}
Spiegelhalter, D.~J., Best, N.~G., Carlin, B.~P., and Van Der~Linde, A. (2002).
\newblock Bayesian measures of model complexity and fit.
\newblock {\em Journal of the royal statistical society: Series b (statistical
  methodology)}, 64(4):583--639.

\bibitem[Talbert and Potoski, 2002]{talbert2002setting}
Talbert, J.~C. and Potoski, M. (2002).
\newblock Setting the legislative agenda: The dimensional structure of bill
  cosponsoring and floor voting.
\newblock {\em Journal of Politics}, 64(3):864--891.

\bibitem[Tsai, 2020]{tsai2020influence}
Tsai, T.-h. (2020).
\newblock The influence of the president and government coalition on roll-call
  voting in brazil, 2003--2006.
\newblock {\em Political Studies Review}, page 1478929920904588.

\bibitem[Watanabe, 2013]{watanabe2013waic}
Watanabe, S. (2013).
\newblock Waic and wbic are information criteria for singular statistical model
  evaluation.
\newblock In {\em Proceedings of the Workshop on Information Theoretic Methods
  in Science and Engineering}, pages 90--94.

\bibitem[Yu and Rodriguez, 2019a]{yu2019circle}
Yu, X. and Rodriguez, A. (2019a).
\newblock Spatial voting models in circular spaces: A case study of the u.s.
  house of representatives.
\newblock {\em Available at SSRN 3381925}.

\bibitem[Yu and Rodriguez, 2019b]{yu2019spherical}
Yu, X. and Rodriguez, A. (2019b).
\newblock Spherical latent factor model.
\newblock {\em Available at SSRN 3381925}.

\bibitem[Zellner, 1986]{zellner1986assessing}
Zellner, A. (1986).
\newblock On assessing prior distributions and bayesian regression analysis
  with g-prior distributions.
\newblock {\em Bayesian inference and decision techniques}.

\bibitem[Zucco, 2013]{zucco2013legislative}
Zucco, C. (2013).
\newblock Legislative coalitions in presidential systems: the case of uruguay.
\newblock {\em Latin American politics and society}, 55(1):96--118.

\bibitem[Zucco and Lauderdale, 2011]{zucco2011distinguishing}
Zucco, C. and Lauderdale, B.~E. (2011).
\newblock Distinguishing between influences on brazilian legislative behavior.
\newblock {\em Legislative Studies Quarterly}, 36(3):363--396.

\end{thebibliography}
\bibliographystyle{apalike}

\appendix

\section{MCMC Algorithm}\label{appendix}

In the same spirit of \cite{albert1993bayesian}, we rely on the fact that any logit or probit model can be expressed as a latent linear regression model. In the case of a probit model, we have that:
\begin{equation*} \label{eqn:regLatente}
    y_{i,j}^{*} = \mu_{j}+\boldsymbol{\alpha}_{j}^{\textsf{T}}\boldsymbol{\beta}_{i}+\epsilon_{i,j}\,, \hspace{0.5cm} \hspace{0.5cm} \epsilon_{i,j} \stackrel{\text {iid}}{\sim}{\text{N}}(0,1)\, .
\end{equation*}
The corresponding expression in the logit case is similar to the previous one, except that the $\epsilon_{i,j}$ follow a standard Logistic distribution  instead of a standard Normal distribution. 
Thus, the parameter space is increased by including auxiliary variables $y_{i,j}^*$, which makes it easier to sample the $\boldsymbol{\beta}_i$, $\boldsymbol{\alpha}_j$, and $\mu_j$. 

The steps of the algorithm are detailed below ($b$ indexes iterations).
\begin{enumerate}
    \item $y_{i,j}^{*(b)}$ is sampled from the full conditional distribution
    $p(y_{i,j}^{*} \mid y_{i,j},\mu_{j}, \boldsymbol{\alpha}_{j}, \boldsymbol{\beta}_{i})$
    as follows:
    \begin{enumerate}
        \item Evaluate $\rho_{i,j}^{(b-1)}=\mu_{j}^{(b-1)}+\boldsymbol{\alpha}_{j}^{(b-1)\,\textsf{T}}\boldsymbol{\beta}_{i}^{(b-1)}$.
        \item Sample $y_{i,j}^{*}$ from a Truncated Normal distribution depending on the observed value $y_{i,j}$ as follows:
            $$p(y_{i,j}^{*} \mid \mu_{j}^{(b-1)},\boldsymbol{\alpha}_{j}^{(b-1)},\boldsymbol{\beta}_{i}^{(b-1)})=\left\{\begin{matrix}
            \text{N}_{(0,\infty)}(y_{i,j}^{*} \mid \rho_{i,j}^{(b-1)},1) & \text{if} & y_{i,j}^{*}=1\,,\\ \\
            \text{N}_{(-\infty,0]}(y_{i,j}^{*} \mid \rho_{i,j}^{(b-1)},1) & \text{if} & y_{i,j}^{*}=0\,.
            \end{matrix}\right.$$
       
    \end{enumerate}
    
    \item $\mu_{j}^{(b)}$ and $\boldsymbol{\alpha}_{j}^{(b)}$ are sampled from the full conditional distribution
    $p(\mu_{j},\boldsymbol{\alpha}_{j} \mid \boldsymbol{B}, y_{i,j}^{*})$
    as follows:
    \begin{enumerate}
        \item Compute $\boldsymbol{c}_{j}$ and $\boldsymbol{C}$, with
        $$\boldsymbol{c}_{j}=[\boldsymbol{B}^{*\textsf{T}}\boldsymbol{B}^{*}+\boldsymbol{A}_{0}^{-1}]^{-1}[\boldsymbol{B}^{*\textsf{T}}\boldsymbol{y}_{.j}^{*(t)}+ \boldsymbol{A}_{0}^{-1}\boldsymbol{a}_{0}] \quad\text{and}\quad \boldsymbol{C}=[\boldsymbol{B}^{*\textsf{T}}\boldsymbol{B}^{*}+\boldsymbol{A}_{0}^{-1}]^{-1}$$
        where $\boldsymbol{B^{*}}$ is an array of $n \times (d + 1)$ whose $i$-th row is $\boldsymbol{\beta}_{i}^{*}=(1, \boldsymbol{\beta}_{i}^{(b-1)})$, and $\boldsymbol{y}_{. j}^{*(b)}$ is a vector of $n \times 1$ storing samples of the latent dependent variables for the $j$-th proposal.
        \item Sample $(\mu_{j}^{(b)},\boldsymbol{\alpha}_{j}^{(b)}) \sim \text{N}(\boldsymbol{c}_{j},\boldsymbol{C})$.
    \end{enumerate} 
    
     \item Note the latent linear regression model is rewritten as
     $w_{i,j}=y_{i,j}^{*}-\mu_{j}=\boldsymbol{\alpha}_{j}^{\textsf{T}}\boldsymbol{\beta}_{i}+\epsilon_{i,j}$. 
     If $\boldsymbol{w}_{i}=(w_{i,1}, \cdots, w_{i, m})$ is taken as the vector of observations associated with legislator $i$, and $\boldsymbol{A} = [\boldsymbol{\alpha}_1,\ldots,\boldsymbol{\alpha}_m]^{\textsf{T}}$ as the corresponding design matrix, the previous equations can be expressed as $\boldsymbol{w}_{i}=\boldsymbol{A}\boldsymbol{\beta}_{i}$. Thus, $\boldsymbol{\beta}_{i}^{(b)}$ is sampled from the full conditional distribution
     $p(\boldsymbol{\beta}_{i} \mid \mu_{j}, \boldsymbol{\alpha}_{j}, y_{i,j}^{*})$ as follows:
     \begin{enumerate}
        \item Compute $\boldsymbol{h}_{i}$ and $\boldsymbol{H}_{i}$, with
        $$\boldsymbol{h}_{i} =[\boldsymbol{A}^{\textsf{T}}\boldsymbol{A}+\boldsymbol{B}_{i}^{-1}]^{-1}[\boldsymbol{A}^{\textsf{T}}\boldsymbol{w}_{i}+ \boldsymbol{B}_{i}^{-1}\boldsymbol{b}_{i}] \quad\text{and}\quad \boldsymbol{H}_{i} =[\boldsymbol{A}^{\textsf{T}}\boldsymbol{A}+\boldsymbol{B}_{i}^{-1}]^{-1}\,.$$
        \item Sample $\boldsymbol{\beta}_{i}^{(b)}\sim \text{N} (\boldsymbol{h}_{i}, \boldsymbol{H}_{i}) \,.$
      \end{enumerate}
     
\end{enumerate}

\end{document}